\documentclass[letter,a4paper]{aa} 
\usepackage{amsmath}
\usepackage[varg]{txfonts}
\usepackage{graphicx}
\usepackage[a4paper]{geometry}
\usepackage{natbib}

\def\ltsima{$\; \buildrel < \over \sim \;$}
\def\simlt{\lower.5ex\hbox{\ltsima}}
\def\gtsima{$\; \buildrel > \over \sim \;$}
\def\simgt{\lower.5ex\hbox{\gtsima}}

\def\arcsec{\hbox{$^{\prime\prime}$}}

\begin{document}

\title{First results from the CALYPSO IRAM-PdBI survey\thanks{Based on observations carried out with the IRAM Plateau de Bure Interferometer. IRAM is supported by INSU/CNRS (France), MPG (Germany), and IGN (Spain).}}
\subtitle{II. Resolving the hot corino in the Class 0 protostar NGC~1333-IRAS2A}
\titlerunning{Resolving the hot corino in the Class 0 protostar NGC~1333-IRAS2A}

\author{A. J. Maury\inst{1,2,4}
\and A. Belloche\inst{3}
\and Ph. Andr\'e\inst{4}
\and S. Maret\inst{5}
\and F. Gueth\inst{6}
\and C. Codella\inst{7}
\and S. Cabrit\inst{8,5}
\and L. Testi\inst{2,7,9}
\and S. Bontemps\inst{10,11}
}

\institute{Harvard-Smithsonian Center for Astrophysics, 60 Garden street, Cambridge, MA 02138, USA
\and ESO, Karl Schwarzschild Strasse 2, 85748 Garching bei M\"unchen, Germany
\and Max-Planck-Institut f\"ur Radioastronomie, Auf dem H\"ugel 69, 53121 Bonn, Germany
\and Laboratoire AIM-Paris-Saclay, CEA/DSM/Irfu - CNRS - Universit\'e Paris Diderot, CE-Saclay, F-91191 Gif-sur-Yvette, France
\and UJF-Grenoble1/CNRS-INSU, Institut de Plan\'etologie et d'Astrophysique de Grenoble, UMR 5274, Grenoble 38041, France
\and IRAM, 300 rue de la Piscine, 38406 St Martin d'H\`eres, France
\and INAF-Osservatorio Astrofisico di Arcetri, Largo E. Fermi 5, I-50125 Firenze, Italy
\and LERMA, CNRS UMR 8112, Observatoire de Paris, ENS, UPMC, UCP, PSL, F-75014 Paris
\and Excellence Cluster Universe, Boltzmannstr. 2, D-85748, Garching, Germany
\and Universit\'e de Bordeaux, LAB, UMR 5804, F-33270 Floirac, France
\and CNRS, LAB, UMR 5804, F-33270 Floirac, France
}

\date{Received Nov. 12, 2013 / Accepted Jan. 22, 2014}

\abstract
{}
{We investigate the origin of complex organic molecules (COMs) in the gas phase around the low-mass Class~0 protostar NGC1333-IRAS2A, to determine if the COM emission lines trace an embedded disk, shocks from the protostellar jet, or the warm inner parts of the protostellar envelope.}
{In the framework of the CALYPSO\thanks{CALYPSO is the Continuum And Lines in Young ProtoStellar Objects survey.} IRAM Plateau de Bure survey, we obtained large bandwidth spectra at sub-arcsecond resolution towards NGC~1333-IRAS2A. We identify the emission lines  towards the central protostar and perform Gaussian fits to constrain the size of the emitting region for each of these lines, tracing various physical conditions and scales.}
{The emission of numerous COMs such as methanol, ethylene glycol, and methyl formate is spatially resolved by our observations. This allows us to measure, for the first time, the size of the COM emission inside the protostellar envelope, finding that it originates from a region of radius 40--100~AU, centered on the NGC~1333-IRAS2A protostellar object. Our analysis shows no preferential elongation of the COM emission along the jet axis, and therefore does not support the hypothesis that COM emission arises from shocked envelope material at the base of the jet. Down to similar sizes, the dust continuum emission is well reproduced with a single power-law envelope model, and therefore does not favor the hypothesis that COM emission arises from the thermal sublimation of grains embedded in a circumstellar disk. Finally, the typical scale $\sim$60~AU observed for COM emission is consistent with the size of the inner envelope where $T_{\rm{dust}} > 100$~K is expected. Our data therefore strongly suggest that the COM emission traces the hot corino in IRAS2A, i.e., the warm inner envelope material where the icy mantles of dust grains evaporate because they are passively heated by the central protostellar object.}
{}

\maketitle

\section{COM emission in Class~0 protostars} 

Along the path leading to the formation of solar-type stars, the Class~0 phase is the main accretion phase during which most of the final stellar mass is accreted onto the central protostellar object \citep{Andre00}.
It is therefore of paramount importance to study the properties of the infalling envelope material on all scales during the Class 0 phase. Ultimately, this will allow us to constrain the efficiency of the accretion/ejection process to build solar-type stars and to shed light on the initial conditions for the formation of protoplanetary disks and planets around stars like our own.
However, on the small scales ($\simlt$50--200~AU) where protostellar disks (progenitors of the protoplanetary disks observed at later stages) are assembled during the embedded phases, most envelope tracers (e.g., $^{12}$CO, N$_{2}$H$^{+}$) are optically thick or chemically destroyed, making it difficult to probe the physical properties of the inner envelope on scales where accretion actually proceeds.

The origin of complex organic molecule (COM) emission in low-mass Class~0 protostars is still debated \citep{Bottinelli07}. First, it has been suggested that the emission of COMs comes from hot corino regions deeply embedded in Class~0 protostars, analogous to the hot cores observed towards high-mass protostars \citep[e.g.,][]{vanDishoeck98, Ceccarelli04, Bottinelli04a}. 
In its early stages, the central protostellar object radiatively heats the surrounding inner envelope. When the temperature of the envelope material becomes high enough ($\sim$100 K), the icy mantles of dust grains evaporate, leading to high gas-phase abundances of complex organic molecules that are formed on dust grains \citep{Cazaux03, Ceccarelli07, Garrod08}. Because of the low luminosities of the central protostars, the region where these molecules are released is expected to be small (typically $\simlt$200 AU in diameter;\citealt{Bottinelli04b}).  
If the COM emission indeed traces the warm inner envelope heated by the central protostellar object, COM lines could be good candidates to study the infall and accretion of circumstellar material down to the very vicinity of the protostellar object itself.
However, it has also been proposed that COM emission could originate from the warm surface of embedded disks \citep{Jorgensen05a}, or shocked material along protostellar jets, since species such as CH$_{3}$OH are known to be greatly enhanced in shocks \citep{Bachiller97b, Chandler05} as a result of grain mantle sputtering.

The CALYPSO survey\footnote{See {\url{http://irfu.cea.fr/Projets/Calypso/}}}, carried out with the IRAM Plateau de Bure (PdBI) interferometer, is providing us with detailed, extensive observations of 17 Class 0 protostars between 92\,GHz and 232\,GHz. One of the main goals of this ambitious observing program is to understand how the circumstellar envelope is being accreted onto the central protostellar object during the Class~0 phase. 
Here, we show that our broad band CALYPSO observations of the Class~0 protostar NGC~1333-IRAS2A ($d\sim235$~pc; \citealt{Hirota08}) provide direct imaging of numerous emission lines, spatially resolving the COM emission structure and therefore putting strong constraints on its origin.

\section{Observations and data reduction}

The Class~0 protostar NGC~1333-IRAS2A (hereafter IRAS2A, see also \citealt{Jennings87, Looney00}) was observed with the IRAM-PdBI, at 218.5~GHz, using the WideX backends to cover the full 3.8\,GHz spectral window at low spectral resolution. Higher resolution backends were placed onto a handful of molecular emission lines: two letters present the analysis of these data, used to explore the kinematics in the inner envelope (with methanol lines, Maret et al.) and the jet properties from molecular line emission (with SO and SiO lines, Codella et al.).
A-array observations were obtained during two observing sessions, respectively in January and February 2011, while C-array observations were carried out in December 2010. The baselines sampled in our observations range from 19~m to 762~m, allowing us to recover emission on scales from $\sim 8\arcsec$ down to $0\farcs35$. Calibration was carried out following standard procedures, using CLIC which is part of the GILDAS\footnote{{\url{http://www.iram.fr/IRAMFR/GILDAS}}} software.
For both A tracks, phase was stable (rms $<$50$^{\circ}$) and pwv was 0.5--1~mm with system temperatures $\sim$100-160\,K, leading to less than 30\% flagging in the dataset.
For the C track, phase rms was $<$80$^{\circ}$, pwv was 1--2~mm, and system temperatures were $\sim$150-250\,K, leading to less than 10\% flagging in the resulting uv-table.  

We merged the datasets to create the final visibility table for the spectra towards IRAS2A. The continuum was built by using the line-free channels, then continuum visibilities were subtracted from the dataset to create a pure spectral cube covering frequencies between 216.85~GHz and 220.45~GHz with a spectral resolution of 3.9\,Mhz ($\sim$2.7\, km s$^{-1}$).
Using natural weighting, the synthesized FWHM beam is $\sim 0\farcs8 \times 0\farcs7$, with an rms noise of $\sim$3~mJy/beam in the line-free channels of the spectra. 

The line identification also used two additional CALYPSO PdBI datasets, consisting of wideband spectra around 231~GHz (setup S1) and 93~GHz (setup S3). These observations are not used in our analysis of the spatial distribution of the COM emission presented in Sect. 3 because their spatial resolution is lower than the spectra analyzed here (setup S2). These datasets will be described in a forthcoming paper that analyzes the chemistry in the IRAS2A envelope.

\section{Line identification}
 
We extracted the WideX spectrum at the continuum emission peak ($\alpha_{\rm{J2000}}=$ 03$^{\rm{h}}$28$^{\rm{m}}$55.575$^{\rm{s}}$, $\delta_{\rm{J2000}}=$ 31$^{\circ}$14$\arcmin$37.05$\arcsec$). The spectrum, presented in Fig.\,\ref{fig:widex_spectrum}, shows a wealth of emission lines whose properties are analyzed in the following.
The method used to identify the detected lines is  described in Appendix A.
We report the first tentative detection of the vibrationally excited state 
$\varv_5 = 1$ of HNCO in a low-mass protostar (2 lines detected in setup S2, see Table\,\ref{t:n1333-ir2a_linelist_s2}). The small number of detected lines prevents us from claiming a firm detection, but these identifications are very plausible since the intensities of the lines are consistent with the model that fits the 
vibrational ground state emission of HNCO.
Moreover, our data provides the first interferometric detection, in a Class 0 protostar, of deuterated methanol (CH$_{2}$DOH: 4 lines detected in setup S2 and 14 lines detected in setups S1/S3 ), ethylene glycol ($aGg'$-(CH$_2$OH)$_2$: 5 lines detected in setup S2 and 11 lines detected in setups S1/S3), and a tentative detection for formamide (NH$_{2}$CHO: 1 line in setup S1 and 1 line in setup S2).

\section{Modeling of the visibilities}

\begin{figure*}
\centering
\includegraphics[width=0.98\linewidth]{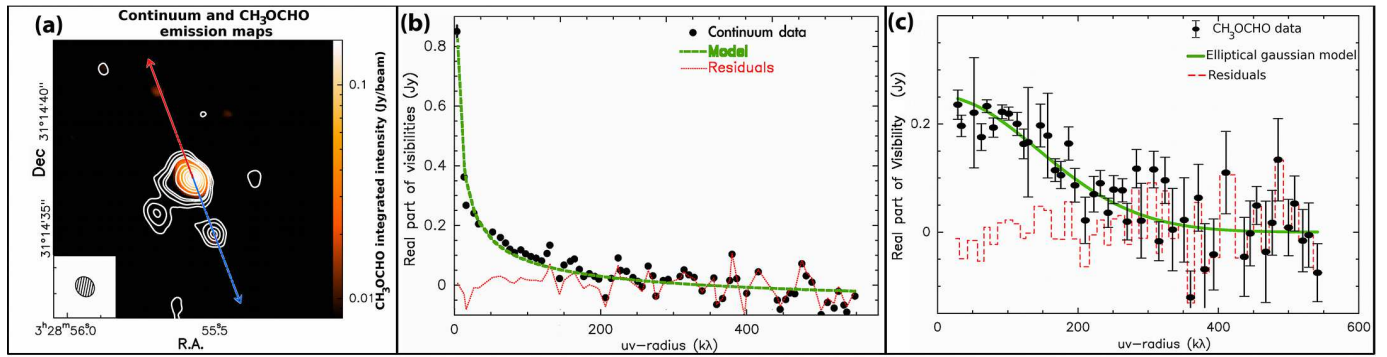}
\caption{
\small{{\bf{(a)}} CH$_{3}$OCHO at 216.966\,GHz (background image) and continuum around  219\,GHz (contours) emission maps towards IRAS2A. The rms noise level is $\sigma = 1.5$ mJy/beam in the continuum, and $\sigma = 2.8$ mJy/beam in the CH$_{3}$OCHO map (emission from one 2.7\,km s$^{-1}$ - channel). Contours are levels at 3$\sigma$, 5$\sigma$, 8$\sigma$ and then from 10$\sigma$ to 100$\sigma$ in 10$\sigma$ increment. The synthesized beam is $0\farcs8 \times 0\farcs68$ (P.A. $32^{\circ}$). The red and blue arrows show the direction of the main bipolar jet axis \citep{Jorgensen07b}. 
{\bf{(b)}} Continuum (real part of the visibilities, black points) and corresponding best-fit power-law model (green dashed line) averaged over baseline bins of 10\,m, as a function of baseline length. Residuals are shown as a red line. The zero-spacing flux is extrapolated from the single-dish flux at 850\,$\mu$m by \citet{Motte01a}.
{\bf{(c)}}: Real part of the visibilities (black points) from the CH$_{3}$OCHO (20 0 20 2 - 19 1 19 1) emission line and corresponding best-fit Gaussian model (FWHM $0\farcs47\times 0\farcs40$, P.A 41$^{\circ}$, green line). Residuals are shown as a red line.
}}
\label{fig:ch3ocho_cont_fit}
 \end{figure*}

The PdBI 1.4 mm dust continuum emission map is shown as contours in Fig.\,\ref{fig:ch3ocho_cont_fit}(a). 
The present analysis only focuses on the emission associated with IRAS2A (MM1); the two secondary continuum sources seen at $[1\farcs53 ,   -1\farcs49]$ (MM2, 15 mJy/beam) and $[-0\farcs92 , -2\farcs21]$ (MM3, 21 mJy/beam) were fitted as compact sources and removed from the continuum visibility table. 

The resulting real part of the continuum visibility of the central source is plotted in Fig.\,\ref{fig:ch3ocho_cont_fit}(b) as a function of uv radius. 
The flux on baselines $\sim200\, k\lambda$ is only half that predicted by interpolation of previous data on the same baselines at 90 GHz and 345 GHz \citep{Looney03, Jorgensen05c}; this difference could be due to secondary sources offset from the phase center, which were not removed (or well resolved) in earlier studies and introduced a positive bias in the visibility amplitudes used by these authors. Figure\,\ref{fig:ch3ocho_cont_fit}(b) also reveals that the real part of the continuum visibility of the central IRAS2A source exhibits a smooth decline all the way out to our longest baseline of $\simeq 550\, k\lambda$, with no sign of residual positive flux on resolved scales. We therefore performed a simple power-law fit to the continuum visibilities of the central source, shown in Fig.\,\ref{fig:ch3ocho_cont_fit}(b). 
In the Rayleigh-Jeans approximation and for optically-thin dust emission, if the temperature and density in the envelope follow simple radial power laws $\rho \propto r^{-p}$ and $T \propto r^{-q}$, the emergent dust continuum emission also has a simple power-law form 
$I(r) \propto r^{-(p+q-1)}$. 
For interferometric observations, the visibility distribution is $V(b) \propto b^{(p+q-3)}$, solely determined by the power-law indices of the temperature and density profiles in the envelope. 
We find that a power-law function $V(b) \propto b^{(-0.45\pm0.05)}$, shown as a green line in Fig.\,\ref{fig:ch3ocho_cont_fit}(b), reproduces well both the continuum emission visibility profile obtained from PdBI and the single-dish flux (0.85 Jy; \citealt{Motte01a}). 
The continuum emission down to r$\sim$35\,AU can therefore be reproduced by a protostellar envelope model without the need for an additional large (200--300 AU diameter) disk component, previously suggested by \citet{Jorgensen05c}.
The residuals map shown in electronic Fig.\,\ref{fig:res_maps}(a) indeed shows that only the asymmetry on the eastern side of the envelope was not fitted properly by our power-law model. 
We note that our $(p+q)$ value ($\sim 2.55$) is in agreement with the protostellar envelope profile of IRAS2A on larger scales ($p+q \sim 2.6$) proposed from BIMA 2.7\,mm observations \citep{Looney03}. 

Table 1 lists the spatial extent of the emission of each identified line that is detected with a medium or high signal-to-noise ratio towards IRAS2A. 
Owing to excitation and abundance variations, the line emission is unlikely to follow a power-law profile similar to the dust
continuum emission. Therefore we modeled the visibilities
with elliptical Gaussians and point sources to determine the emitting size of each of these molecular lines.
We note that most of the lines are not spectrally resolved: the fit was performed on the emission in the channel showing the greatest flux to avoid possible contamination from neighboring lines; in most cases this is the channel at the systemic velocity of the core. 
An example of this modeling is shown in Fig.\,\ref{fig:ch3ocho_cont_fit}(c).
The best fit was kept when a chi-square minimization converged, complemented by a visual inspection of the residuals maps (see electronic Fig.\,\ref{fig:res_maps}(b) for an example) to ensure that no significant emission was left around the IRAS2A source. The parameters of the best-fit model for each identified molecular line are given in Table\,\ref{t:n1333-ir2a_lineprops_s2}. For the lines whose emission morphology is dominated by outflow- or jet-like structures (SiO and SO mainly), no satisfactory fit of their complex spatial distributions was found with elliptical Gaussian models, and values are not reported.

\section{Discussion: origin of the COM emission}

It is still being debated whether COM emission lines observed in low-mass Class 0 protostars are tracing embedded disks, shocks from protostellar jets, or the warm inner parts of protostellar envelopes.
\cite{Bottinelli04a} observed the Class~0 protostar IRAS\,16293 (I16293 in the following) with the PdBI and found that the emission of CH$_{3}$CN and CH$_{3}$OCHO was  unresolved on $180$\,AU scales, while observations of the NGC1333-IRAS4A protostar showed that the CH$_{3}$CN emission comes from a region $\sim 175$\,AU (FWHM) in diameter \citep{Bottinelli08}. These observations also showed that emission from CH$_{3}$CN and CH$_{3}$OCHO is not detected on large scales along the outflow axis but only on small scales around the embedded protostar. However, \citet{Jorgensen07b} used the Submillimeter Array (SMA) to observe the IRAS4A protostar and showed that CH$_{3}$OH and H$_{2}$CO emission is extended along the outflow axis; and SMA observations of I16293 detected a wealth of COM emission lines \citep{Bisschop08, Jorgensen11}, mostly unresolved. \citet{Pineda12} and \citet{Zapata13} used ALMA observations of I16293 to analyze the spatial and velocity distribution of 3 COM lines, and trace the infall on small scales around source B.
Observations of IRAS2A with an angular resolution of $\sim 2 \arcsec$ detected emission lines from five COMs, all spatially unresolved \citep{Jorgensen05c}. \citet{Persson12} and \citet{Kristensen12} detected water emission in the IRAS2A envelope and outflow, both at the systemic and blue-shifted velocities.

\begin{figure}  \centering
\includegraphics[width=\linewidth]{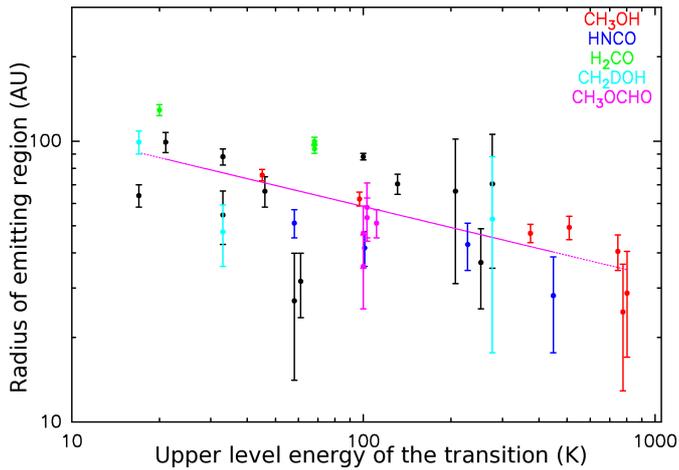}
\caption{\small{Emission size (average radius of the best-fit elliptical Gaussian) of the molecular lines listed in Table\,\ref{t:n1333-ir2a_lineprops_s2} as a function of their upper-level energy. Only the spatially resolved, unblended lines for which a good fit could be achieved are represented. Black symbols show the molecules for which only one emission line is observed and/or resolved in setup S2, while colored symbols show molecules for which several lines are detected and resolved. The error bars are the largest error bar (minor or major axis) produced by the fit with an elliptical Gaussian model. The pink line shows the best power-law fit to the data, as described in the text. All COM lines that are resolved emit in a region of radius $\simlt$100~AU.}
}
\label{fig:Line_prop_correl}
 \end{figure}

Thanks to the bandwidth, sensitivity, and resolution of PdBI, emission from large samples of COMs can now be mapped in the very inner protostellar envelope: our analysis shows that tens of high excitation (E$_{\rm up} \simgt 100$\,K) emission lines are detected and spatially resolved in our PdBI data. They probe various physical conditions (excitation temperatures range from $\sim$30\,K to 250~K for CH$_{3}$OH and HNCO) and a wide range of upper-level energies, therefore allowing us to spatially resolve the hot corino emission in IRAS2A, for the first time.
The emitting sizes of the molecules detected towards IRAS2A are shown in Fig.\,\ref{fig:Line_prop_correl}. All COMs are observed in the inner envelope at radii $\simlt$100\,AU.
The best fit reproducing the dependency of emitting sizes with upper-level energies of the molecular transitions is a power law FWHM $\propto E_{\rm{up}}^ {-0.25\pm0.05}$. However, the exact slope of this correlation is not very tightly constrained because of the limited number of high signal-to-noise lines, and a fit with a constant radius of ~55 AU yields a only marginally higher $\chi^{2}$. The exact dependence of the emitting region radius with $E_{\rm{up}}$ will
have to be explored with more COM lines detected with a higher signal-to-noise ratio.
However, we note that the highest excitation lines that we spatially resolve are all located inside a region of radius $r \sim$40~AU (average FWHM$\sim 0.34\arcsec$ for transitions with 300\,K$<E_{\rm{up}}<$800\,K). 

Modeling the PdBI continuum data, we find that the dust continuum emission is well described by a protostellar envelope with a single power-law profile 
$I(r) \propto r^{-1.55}$ (see Sect.\,3 and Fig.\,\ref{fig:ch3ocho_cont_fit}b), down to physical sizes ($r\sim 35$ AU) probed by our longest baselines. 
This suggests that, on scales where the COMs emit, the dust emission is still predominantly associated to the envelope and not to a protostellar disk. Since most COMs are released from dust grains, it suggests that the COMs are released in the warm inner envelope around the central protostellar object. 
Moreover, the observed size of $r$ 40-100~AU for the COM emission in IRAS2A is in good agreement with predictions from simple analytical models computing the size of the hot corino region at $T\simgt 100$ K: if we assume a density profile $\rho \propto r^{-1.5}$ for the envelope, and with $(p+q-1)=1.55$ (see Sect.\,3), the radius at which dust in the envelope reaches a temperature of 100\,K is $r_{\rm{100K}}\sim60\pm5$ AU ( $40\pm3$ AU with $\rho \propto r^{-2}$, see Eq. 2 in \citealt{Motte01a}). We note, however, that the presence of an unresolved $r <$40 AU disk in IRAS2A is not excluded by our observations, and could contribute to the emission of COMs on the smaller, unresolved scales.

The emission of the least complex molecules (e.g., H$_{2}$CO, DCN, OCS) tends to be elongated along position angles $\simlt 30^{\circ}$ of the protostellar jet P.A - traced by SiO emission (see \citealt{Codella14}). This emission cannot be due to sputtering in shocks since the observed velocities are very close to the systemic velocity of the core. Therefore, these emission lines are most probably due to thermal or photodesorption in the UV-heated cavity walls \citep{Visser12}: this might illustrate the contribution of the irradiated outflow walls, e.g., the warm envelope material with non-spherical geometry at $T_{\rm{dust}} >$ 100\,K, to the hot corino emission. However, we stress that the spatial distribution of the COM emission at high $E_{\rm{up}}$ does not show a preferential elongation along the outflow/jet axis : the median aspect ratio of the best-fit elliptical Gaussian models is $\sim$1.7, but the position angles are very widely distributed from $-80^\circ$ to $85^\circ$ (see Table\,\ref{t:n1333-ir2a_lineprops_s2}).

The spectral resolution of the WideX data prevents us from carrying out any kinematic study of those lines. However, these COM emission lines are detected at velocities close to the systemic velocity of the protostar ($\pm$1.4 km/s, see Table~2). Therefore it is very unlikely that these molecules are formed from non-thermal desorption mechanisms due to the interaction of a collimated high-velocity jet with the surrounding envelope material, as suggested by \cite{Oberg11}. Instead the kinematical and spatial distribution information from our data draw a picture where the COM emission originates from warm material surrounding the protostar and dynamically associated to the protostellar envelope: the so-called hot corino region.
This opens up the perspective of using complex molecules as tracers of the inner envelope kinematics, down to the $\simlt$100~AU sizes where rotationally-supported protostellar disks may form, and accretion proceeds on the protostar itself. In a related paper, we make use of the high spectral resolution observations of two methanol emission lines to assess their kinematical properties \citep{Maret14}. 

To conclude, our results strongly support a scenario where the COM emission in IRAS2A originates from a hot corino, i.e., the release of complex molecules in the gas phase of the inner ($\simlt 100$\,AU) envelope, when a critical temperature allowing sublimation of icy grain mantles is reached, and not from shocks at the base of the jet. Analysis of the CALYPSO observations toward the remaining 16 Class~0 objects in our sample will enhance the statistics on the occurrence of COM emission in Class~0 protostars, and determine if they always trace a hot corino in the warm inner envelope of these objects. If this is the case, the analysis of spectral emission lines from complex molecules will be a powerful tool for tracing the protostellar kinematics during the disk formation epoch, on scales where protoplanetary disks are observed at later (e.g., T-Tauri) stages.

\begin{acknowledgements} We thank Holger M{\"u}ller for the spectroscopic predictions of HNCO,$\varv_5$=1.
We are very thankful to the IRAM staff, whose dedication allowed us to carry out the CALYPSO project. 
This research has received funding from the European Community's Seventh Framework Programme (/FP7/2007-2013/) under grant agreements No 229517 (ESO COFUND) and No 291294 (ORISTARS). S. M. acknowledges the support of the French Agence Nationale de la Recherche (ANR), under reference ANR-12-JS05-0005.
\end{acknowledgements}

\bibliographystyle{aa}
\bibliography{bibliographie}

\begin{appendix}

\section{Line identification}

\begin{figure*}[!t]  
\centering
\includegraphics[width=0.90\linewidth]{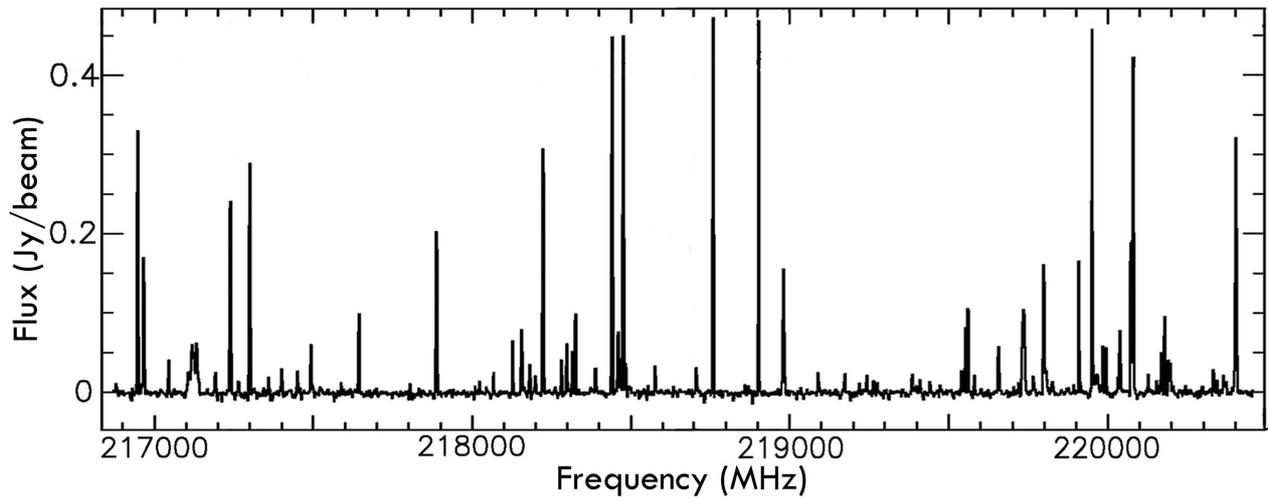}
\caption{Continuum-subtracted WideX spectrum around 218.5\,GHz (setup S2), at the position of the maximum of 1.4 mm continuum emission towards IRAS2A (shown in Fig.\ref{fig:ch3ocho_cont_fit}a). 
\label{fig:widex_spectrum}}
 \end{figure*}
 
We extracted the WideX spectrum at the continuum emission peak ($\alpha_{\rm{J2000}}=$ 03$^{\rm{h}}$28$^{\rm{m}}$55.575$^{\rm{s}}$, $\delta_{\rm{J2000}}=$ 31$^{\circ}$14$\arcmin$37.05$\arcsec$) shown in Fig.\,\ref{fig:widex_spectrum}.
The line identification was performed with the XCLASS 
software\footnote{See http://www.astro.uni-koeln.de/projects/schilke/XCLASS.}
under the assumption of local thermodynamic equilibrium. 
Our spectroscopic database contains all entries of the CDMS \citep{Muller05} and JPL \citep{Pickett98} catalogs, as well as a few
private entries (for more details, see \citealt{Belloche13}).
The spectra were modeled species by species. For most species, an excitation temperature of 100\,K was assumed for the LTE modeling, while for a few species the detection of several transitions with significantly different upper-level energies allowed us to leave the temperature as a free parameter of our modeling (finding temperatures up to $\sim$250 K for CH$_{3}$OH and HNCO, for example).
For each species, the spectra in all three frequency setups were modeled at once, using five parameters: source 
size, temperature, column density, line width, and velocity offset with 
respect to the systemic velocity of the source. The emission of all transitions was assumed to come from a source of size $0\farcs6$ (average FWHM, see Sect.\,3). The fit optimization was performed by eye. For a few species (SO, SiO, CO), it was necessary to include several velocity components to account for the shape of the detected lines.
A total of 86 emission lines with peak signal-to-noise ratios higher than 3 were detected at the position of the continuum emission peak, in setup S2, among which 55 are identified, see Table\,\ref{t:n1333-ir2a_linelist_s2}.
\end{appendix}
 
 \Online

\onltab{
\begin{table*}
\caption{List of emission lines detected at the peak of the dust continuum emission at 219 GHz towards IRAS2A}
\label{t:n1333-ir2a_linelist_s2}
\begin{tabular}{cccc}
\hline
{Frequency$^{a}$} &  {$S/N ^{b}$} &  {Identification $^{c}$} & {Quantum numbers} \\
{(MHz)} &{} & {} &{}  \\
\hline
216879 & low    &       ? &  \\
216947 & high   &       CH$_3$OH &  5 1 4 0 - 4 2 2 0  \\
 &  & + ? &  \\
216966 & high   &       CH$_3$OCHO & 20 0 20 2     - 19 1 19 1  \\
& & & 20 0 20 0     - 19 1 19 0 \\
& & & 20 1 20 1     - 19 1 19 1 \\
& & & 20 1 20 0     - 19 1 19 0 \\
& & & 20 0 20 2     - 19 0 19 2 \\
& & & 20 0 20 0     - 19 0 19 0 \\
& & & 20 1 20 1     - 19 0 19 2 \\
& & & 20 1 20 0     - 19 0 19 0 \\
217045 & high   &       ? &  \\
217108 & high   &       SiO & 5 0 - 4 0   \\
217118 & high   &       SiO & 5 0 - 4 0  \\
217123 & high   &       SiO & 5 0 - 4 0  \\
217133 & high   &       SiO & 5 0 - 4 0  \\
 & & + $aGg'$-(CH$_2$OH)$_2$ & 21 4 17 0 - 20 4 16 1  \\
217192 & high   &       CH$_3$OCH$_3$ & 22 4 19 3 - 22 3 20 3  \\
 & & & 22 4 19 5     - 22 3 20 5  \\
  & & & 22 4 19 1     - 22 3 20 1  \\
   & & & 22 4 19 0     - 22 3 20 0  \\
217238 & high   &       DCN & 3- 2  \\
217263 & low    &       ? &  \\
217300 & high   &       CH$_3$OH, $\varv_{\rm t}=1$ & 6 1 5 -1  - 7 2 6 -1  \\
217359 & medium &       CH$_2$DOH & 17 4 14 2     - 16 5 12 1  \\
217400 & high   &       ? &  \\
217450 & high   &       CH$_2$DOH  & 18 1 17 2     - 18 2 17 0  \\
& & + $aGg'$-(CH$_2$OH)$_2$ & 24 1 24 0     - 23 1 23 1  \\
& & & 24 0 24 0     - 23 0 23 1  \\
217492 & high   &       ? &  \\
217588 & low    &       $aGg'$-(CH$_2$OH)$_2$ & 21 2 19 1     - 20 2 18 0  \\
217643 & high   &       CH$_3$OH, $\varv_{\rm t}=1$ & 15 6 10 -1    - 16 5 11 -1  \\
 & & & 15 6 9 1      - 16 5 12 1  \\
217804 & low    &       ? &  \\
217887 & high   &       CH$_3$OH &  20 1 19 0     - 20 0 20 0  \\
218022 & low    &       ? &  \\
218067 & high   &       ? &  \\
218127 & high   &       ? &  \\
218156 & high   &       ? &  \\
218181 & high   &       ? &  \\
218199 & medium &       ? &  \\
218222 & high   &       H$_2$CO & 3 0 3         - 2 0 2  \\
218281 & high   &       CH$_3$OCHO & 17 3 14 2     - 16 3 13 2  \\
218298 & high   &       CH$_3$OCHO & 17 3 14 0     - 16 3 13 0  \\
218317 & high   &       CH$_2$DOH & 5 2 4 1       - 5 1 5 1  \\
218325 & high   &       HC$_3$N & 24   - 23  \\
218389 & high   &       ? &  \\
218441 & high   &       CH$_3$OH & 4 2 2 0       - 3 1 2 0  \\
218459 & high   &       NH$_2$CHO & 10 1 9        - 9 1 8  \\
218469 & low    &       ?  &  \\
& & + $aGg'$-(CH$_2$OH)$_2$ & 22 14 8 0     - 21 14 7 1  \\
& & & 22 14 9 0     - 21 14 8 1  \\
218476 & high   &       H$_2$CO & 3 2 2         - 2 2 1  \\
218482 & high   &       ? &  \\
218498 & low    &       ? &  \\
218554 & low    &       ? &  \\
218575 & high   &       $aGg'$-(CH$_2$OH)$_2$ & 22 13 9 0     - 21 13 8 1  \\
 & & & 22 13 10 0    - 21 13 9 1  \\
218708 & high   &       ?  &  \\
& & + $aGg'$-(CH$_2$OH)$_2$ & 22 12 10 0    - 21 12 9 1  \\
& & & 22 12 11 0    - 21 12 10 1  \\
218760 & high   &       H$_2$CO &  3 2 1         - 2 2 0  \\
218861 & low  	  &	  HC$_3$N, $\varv_7=1$ &  24 -1         - 23 1  \\
218903 & high   &       OCS & 18            - 17  \\
218981 & high   &       HNCO & 10 1 10       - 9 1 9  \\
\end{tabular}
\end{table*}
\addtocounter{table}{-1}
\begin{table*}
\caption{Continued}
\label{t:n1333-ir2a_linelist_s2}
\begin{tabular}{cccc}
\hline
{Frequency$^{a}$} &  {$S/N ^{b}$} &  {Identification $^{c}$} & {Quantum numbers} \\
{(MHz)} &{} & {} &{} \\
\hline
219089 & medium &       $aGg'$-(CH$_2$OH)$_2$ & 22 10 13 0    - 21 10 12 1  \\
 & & & 22 10 12 0    - 21 10 11 1  \\
219174 & medium &       HC$_3$N, $\varv_7=1$ & 24 1          - 23 -1  \\
219242 & medium &       ? &  \\
219263 & low    &       ? &  \\
219276 & low    &		SO$_2$ & 22 7 15       - 23 6 18  \\
219385 & medium &       $aGg'$-(CH$_2$OH)$_2$ &  22 9 14 0     - 21 9 13 1  \\
 & & & 22 9 13 0     - 21 9 12 1  \\
219398 & low    &       ? &  \\
219410 & low    &       ? &  \\
219441 & low    &       ? &  \\
219474 & low & ? &  \\
219540 & medium &       HNCO, $\varv_{5}=1$ & 10 1 10 2 10  - 9 1 9 2 9  \\
& & + $aGg'$-(CH$_2$OH)$_2$ &  22 2 21 1     - 21 2 20 0   \\
219551 & high   &       HNCO & 10 4 7        - 9 4 6  \\
& & & 10 4 6        - 9 4 5  \\
& & + CH$_2$DOH &  5 1 5 1       - 4 1 4 1  \\
219560 & high   &       C$^{18}$O & 2             - 1  \\
219580 & low    &       $aGg'$-(CH$_2$OH)$_2$ & 22 1 21 1     - 21 1 20 0  \\
219656 & high   &       HNCO &  10 3 8        - 9 3 7  \\
 & & & 10 3 7        - 9 3 6  \\
219719 & low & ? &  \\
219735 & high   &       HNCO & 10 2 9        - 9 2 8  \\
 & & & 10 2 8        - 9 2 7  \\
219765 & medium &       $aGg'$-(CH$_2$OH)$_2$ & 20 4 16 1     - 19 4 15 0  \\
219798 & high   &       HNCO & 10 0 10       - 9 0 9  \\
219803 & high   &       ? &  \\
& & + $aGg'$-(CH$_2$OH)$_2$ & 22 8 15 0     - 21 8 14 1  \\
219825 & low    &       ?  &  \\
& & + CH$_3$OCHO, $\varv_{\rm t}=1$ & 18 10 8 3     - 17 10 7 3  \\
& & & 18 10 9 3     - 17 10 8 3  \\
219892 & low    &       ? &  \\
219908 & high   &       H$_2$$^{13}$CO & 3 1 2   - 2 1 1  \\
219950 & high   &       SO & 5 6    - 4 5  \\
219965 & medium &       SO & 5 6    - 4 5  \\
219984 & high   &       CH$_3$OH & 25 3 22 0     - 24 4 20 0  \\
219994 & high   &       CH$_3$OH &  23 5 19 0     - 22 6 17 0  \\
220030 & low    &       ? &  \\
& & + CH$_3$OCHO, $\varv_{\rm t}=1$ & 18 9 9 3      - 17 9 8 3  \\
& & & 18 9 10 3     - 17 9 9 3  \\
220038 & high   &       t-HCOOH & 10 0 10       - 9 0 9  \\
220072 & high   &       CH$_2$DOH & 5 1 5 0       - 4 1 4 0  \\
220079 & high   &       CH$_3$OH & 8 0 8 0       - 7 1 6 0   \\
220126 & medium &       ? &  \\
220153 & low    &       HNCO,$\varv_5$=1 & 10 0 10 2 10  - 9 0 9 2 9  \\
220167 & high   &       CH$_3$OCHO & 17 4 13 2     - 16 4 12 2  \\
& & + HNCO, $\varv_5 = 1$ & 10 3 7 2 10   - 9 3 6 2 9  \\
 & & & 10 3 8 2 10   - 9 3 7 2 9  \\
220178 & high   &       CH$_2$CO & 11 1 11       - 10 1 10  \\
220191 & high   &       CH$_3$OCHO & 17 4 13 0     - 16 4 12 0  \\
220195 & medium &       ? &  \\
& & + HNCO, $\varv_5$=1 & 10 2 9 2 10   - 9 2 8 2 9  \\
220296 & low & ? &  \\
220332 & high   &       ? &  \\
220343 & low    &       ? &  \\
220363 & medium &       ? &  \\
220369 & low    &       ? &  \\
220402 & high   &       $^{13}$CO & 2             - 1  \\
& & + CH$_3$OH & 10 -5 5 0     - 11 -4 8 0  \\
& & + $aGg'$-(CH$_2$OH)$_2$ & 22 7 16 0     - 21 7 15 1  \\
\hline
\end{tabular}
\tablefoot{$^{a}$ Observed frequency of the emission line. $^{b}$ Peak signal-to-noise ratio of the (tentatively) detected line. High means $>9$, medium between 9 and 6, and low between 6 and 3. $^{c}$ A question mark means an unidentified line. A molecule name accompanied by a question mark refers to a partially unidentified line:  an identified molecule emits at the frequency of the detected line; however its predicted line intensity is weaker than the observed one and thus there must be contribution from another (unidentified) species. All lines separated by $\simlt$4\,MHz are blended in our WideX data. }
\end{table*}
}

\onltab{ 
\begin{table*} 
\caption{Properties of identified emission lines detected toward the continuum peak emission of IRAS2A around 219~GHz}
\label{t:n1333-ir2a_lineprops_s2}
\begin{tabular}{ccccccc}
\hline
{Rest frequency} 
& {Molecule $^{a}$} 
& {E$_{\rm{up}}$$^{b}$} 
& {Size (FWHM)$^{c}$} 
& {P.A.$^{d}$} 
& {Flux $^{e}$} 
& {n$_{\rm{crit, 100K}}$ $^{f}$}\\
{(MHz)} 
& 
& (K) 
& (arcsec) 
& ($^{\circ}$) 
& (Jy/beam) 
& ({cm$^{-3}$)}\\
\hline
216945.60    &       CH$_3$OH (+?)	 		         & 56		& 0.7$\times$0.4 ($\pm$0.03)				& 30 		& 0.51  & 7.1e6     \\
216966.65    &       CH$_3$OCHO				& 111	&  0.47$\times$0.40 ($\pm$0.05)$^{\star}$ 	& 41 		& 0.25 &   \\
217104.98    &       SiO						& 31		& {-} 									& {-} 		& {-} & 4.7e6  \\
217193.17    &       CH$_3$OCH$_3$			& 253	& 0.48$\times$0.15   ($\pm$0.1) 			& 33		& 0.04  &   \\
217238.53    &       DCN	        					& 21		& 1.13$\times$0.56 ($\pm$0.07)$^{\star}$ 	& 34		& 0.37 &    \\
217299.20    &       CH$_3$OH, $\varv_{\rm t}=1$				& 374	&  0.5$\times$0.3 ($\pm$0.03)$^{\star}$ 		& 41 		& 0.40   &   \\
217359.28    &       CH$_2$DOH				& 277	& 0.9$\times$0.3 ($\pm$0.3) 				& $-$45	& 0.04   &    \\
217447.90    &       CH$_2$DOH (+$aGg'$-(CH$_2$OH)$_2$) 		& 265 	& point  								& {}		&  0.043  &  \\
217642.86    &       CH$_3$OH, $\varv_{\rm t}=1$				& 746	& 0.46$\times$0.23 ($\pm$0.05) 			& 62		& 0.14    &    \\
217886.39    &       CH$_3$OH					& 508	& 0.51$\times$0.33 ($\pm$0.04)$^{\star}$ 	& 48		& 0.30   &   \\
218222.19    &       H$_2$CO	        				&  20	 	& 1.4$\times$0.8 ($\pm$0.05)$^{\star\star}$ 	& 25		& 0.83  & 5.0e6  \\
218280.90    &       CH$_3$OCHO				& 100	& 0.45$\times$0.35 ($\pm$0.1) 			& $-$31	& 0.07   &   \\
218297.89    &       CH$_3$OCHO				& 100	& 0.41$\times$0.20 ($\pm$0.09) 			& 21		& 0.09 &    \\
218316.39    &       CH$_2$DOH				& 33		& 0.64$\times$0.29  ($\pm$0.1) 			& $-$36	& 0.09  &   \\
218324.71    &       HC$_3$N					& 131	& 0.75$\times$0.45 ($\pm$0.05)$^{\star}$	& 21		& 0.19  & \\
218440.05    &       CH$_3$OH					& 45		& 0.79$\times$0.50 ($\pm$0.03)$^{\star}$ 	& 27		& 0.81 & 7.8e7  \\
218459.65    &       NH$_2$CHO				& 61		& 0.34$\times$0.20 ($\pm$0.07)$^{\star\star}$ 	& 72		& 0.1  &   \\
218475.63    &       H$_2$CO	        				& 68		& 1.0$\times$0.60 ($\pm$0.03)$^{\star\star}$ 	& 22		& 1.0 & 5.6e6   \\
218574.68    &       $aGg'$-(CH$_2$OH)$_2$			&  207	& 0.88$\times$0.25 ($\pm$0.3) 			& 40		& 0.05   &  \\
218705.81    &       (?+) $aGg'$-(CH$_2$OH)$_2$	& 195	& point 								& {}		& 0.04  &    \\
218760.07    &       H$_2$CO	        				& 68		& 1.0$\times$0.70 ($\pm$0.03)$^{\star\star}$ 	& 21		& 0.96 & 6.1e6   \\
218903.35    &       OCS	        					& 100	& 0.9$\times$0.6 ($\pm$0.02)$^{\star}$ 		& 25		& 0.89 & 4.0e5   \\
218981.17    &       HNCO	       					& 101	& 0.45$\times$0.26 ($\pm$ 0.05)$^{\star}$ 	& 35		& 0.21 & 8.2e7    \\
219089.73    &       $aGg'$-(CH$_2$OH)$_2$			& 173	& point 								& {}		& 0.04  &    \\
219173.75    &       HC$_3$N, $\varv_7=1$		& 452	& point								& {}		& 0.03	& \\
219385.18    &       $aGg'$-(CH$_2$OH)$_2$ 			& 164	&  point 								& {}		& 0.03   &  \\
219540.33    &       HNCO, $\varv_{5}=1$ (+$aGg'$-(CH$_2$OH)$_2$)	 	& 902		& point 								& {}		& 0.04  &    \\
219547.09    &       HNCO (+ CH$_2$DOH)		& 709	& point 								& {}		& 0.03   &    \\
219551.48    &       CH$_2$DOH (+ HNCO)		& 26		& 0.57$\times$0.26 ($\pm$ 0.09) 			& 61		& 0.12   &   \\
219560.35    &       C$^{18}$O					& 16		& 5.0$\times$4.0 ($\pm$ 0.15) 				& $-$80	& 3.0 & 9.6e3   \\
219580.67    &       $aGg'$-(CH$_2$OH)$_2$			& 122 	&  point 								& {}		& 0.03  &   \\
219656.71    &       HNCO	        					& 448	& 0.38$\times$0.10 ($\pm$ 0.09)$^{\star}$ 	& 23		& 0.08   &    \\
219733.85    &       HNCO	     					& 228	& 0.59$\times$0.14 ($\pm$ 0.07) 			& 39		& 0.15  &    \\
219764.92    &       $aGg'$-(CH$_2$OH)$_2$			& 113	& point 								& {}		& 0.03  &     \\
219798.32    &       HNCO	       					& 58		& 0.60$\times$0.27 ($\pm$0.05)$^{\star}$ 	& 37		& 0.24 & 7.6e6    \\
219803.67    &       (?+) $aGg'$-(CH$_2$OH)$_2$	& 156	& point 								& {}		& 0.05  &   \\
219908.48    &	     H$_2$$^{13}$CO			&  33		& 0.93$\times$0.57($\pm$ 0.05)			& 19		& 0.34 & \\
219949.44    &       SO	        					& 35		&  {-}									& {-} 		& {-} & 3.7e6  \\
219983.99    &       CH$_3$OH					& 802	& 0.31$\times$0.18  ($\pm$0.1) 			& 53		& 0.08  &    \\
219993.94    &       CH$_3$OH					& 776	& 0.31$\times$0.11 ($\pm$0.1) 			& 63		& 0.07  &   \\
220038.07    &       t-HCOOH					& 58 		& 0.39$\times$0.07 ($\pm$0.11)$^{\star}$ 	& 33		& 0.10   & \\
220071.80    &       CH$_2$DOH				& 17		&  0.65$\times$0.44 ($\pm$0.05) 			& 42		& 0.31  &  \\
220078.49    &       CH$_3$OH					& 97		& 0.63$\times$0.43  ($\pm$0.03)$^{\star}$ 	& 25 		& 0.65 & 2.9e7     \\
220166.88    &       CH$_3$OCHO (+	HNCO, $\varv_{5}=1$)		& 103	& 0.71$\times$0.20 ($\pm$0.08) 			& 55		& 0.08 		&   \\
220178.19    &       CH$_2$CO					& 46		& 0.60$\times$0.53 ($\pm$0.07) 			& 85		& 0.16   &   \\
220190.28    &       CH$_3$OCHO				& 103	&  0.72$\times$0.27 ($\pm$0.11) 			& 55		& 0.08   \\
220193.86    &      (?+) HNCO, $\varv_{5}=1$ 		& 967	& 0.60$\times$0.39 ($\pm$0.16) 			& $-$5	& 0.07   &    \\
220398.68    &      $^{13}$CO (+CH$_3$OH) 		& 16     	& 9.5$\times$5.4 ($\pm$ 0.30)  			& 47		& 3.6 & 9.7e3 \\
220401.37    &      CH$_3$OH (+$^{13}$CO)		& 252	& 0.45$\times$0.25   ($\pm$0.04) 			& 48		& 0.28 & 9.3e6   \\
\hline
\end{tabular}
\tablefoot{
$^{a}${Molecule identified to be the source of the line emission. When two lines are too close in frequencies to be separated with the WideX channel width, the secondary spectral line is indicated in parentheses.}\\
$^{b}${Upper-level energy of the transition.}\\
$^{c}${Size of the best fit of an elliptical Gaussian to the visibilities, see text for further details. Gaussian fits were performed on the channel showing the maximum emission (our data has a spectral resolution of 3.9\,MHz $\sim$2.7\, km\,s$^{-1}$). Most lines are detected at the systemic velocity 
$V_{\rm{sys}} \sim 7(\pm0.5)$ km\,s$^{-1}$. Two notable lines show several blue-shifted velocity components: SiO shows emission at 7, 3, -14.5, and -32.5 km\,s$^{-1}$, while SO is detected at velocities 6.5, -0.5, and -14.0 km\,s$^{-1}$. 
No FWHM is indicated when the molecular emission is dominated by an outflow- or jet-like structure, as determined by a preliminary inspection of the emission maps line per line. The $^{\star}$ symbol indicates the presence of a secondary component to be fitted at a different position in the map. The $^{\star}$$^{\star}$ symbol indicates that, in addition to the component towards IRAS2A the line also traces outflow and/or complex structures at other locations in our maps.}\\
$^{d}${Position angle of the elliptical Gaussian, when an elliptical Gaussian fit  could be performed successfully.}\\
$^{e}${Flux attributed to the best fit with a Gaussian component located at the center of the protostellar envelope.}\\
$^{f}${Critical density of the transition, computed at 100\,K, using the collision rates compiled in the LAMDA database when available. \url{http://home.strw.leidenuniv.nl/~moldata/}}
}
 \end{table*}
}

\onlfig{
\begin{figure*} 
\centering
\includegraphics[width=0.5\linewidth]{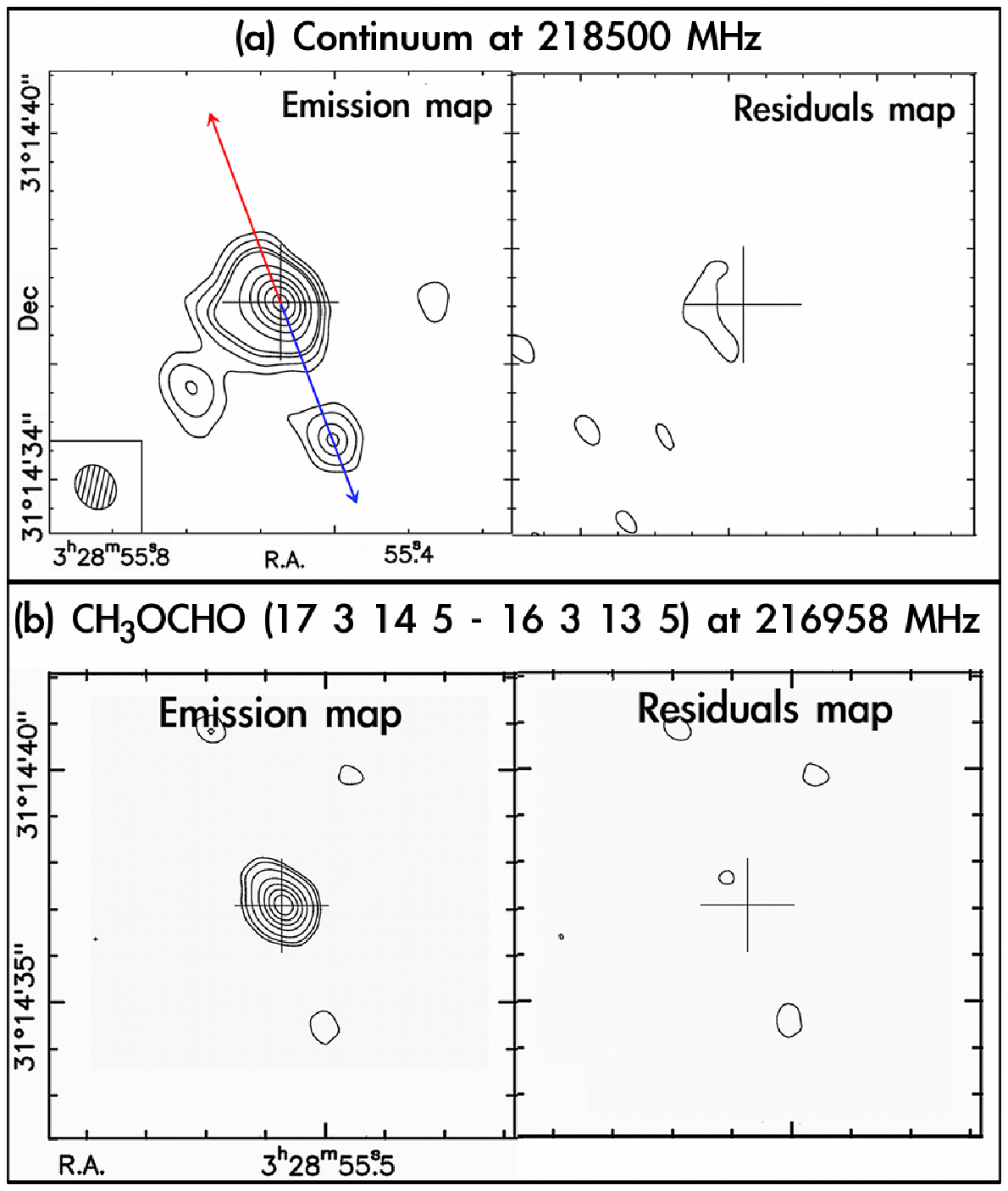}
\caption{\small{Emission and residual maps of the continuum emission and CH$_3$OCHO line emission at 217\,GHz. {\bf{(a)}} The left and right panels, respectively, show the PdBI continuum emission map and residuals map. The residuals map was obtained by removing the two secondary sources as point sources, then removing the best-fit power-law model visibilities from the data visibilities, and imaging the residuals table. Contours show the levels of 3$\sigma$, 5$\sigma$, and 8$\sigma$, and then 10$\sigma$ to 100$\sigma$ in 10$\sigma$ steps. The cross shows the phase center of our observations, coinciding with the peak of the continuum emission at 1.4\,mm.
{\bf{(b)}} The left and right panels, respectively, show the CH$_3$OCHO emission and residual maps. In the maps, the rms noise level is $\sigma = 2.8$ mJy/beam. Contours show the 3$\sigma$, 5$\sigma$, and 8$\sigma$, and then 10$\sigma$ to 60$\sigma$ in 10$\sigma$ steps.}}
\label{fig:res_maps}
 \end{figure*}
 }

\onlfig{
\begin{figure*} 
\centering
\includegraphics[width=0.98\linewidth]{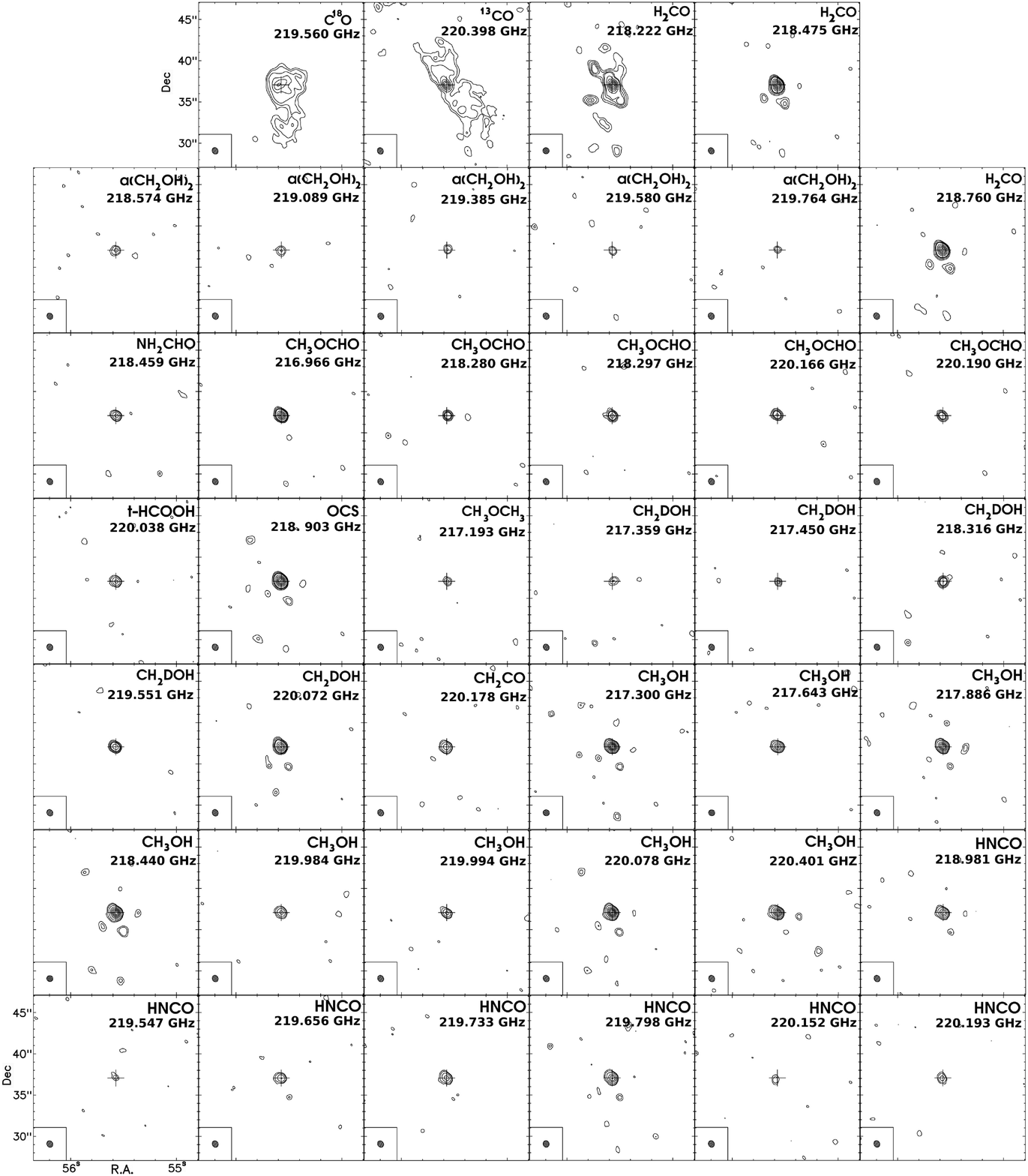}
\caption{\small{Emission maps for most of the identified molecular emission lines in the spectrum of IRAS2A. While CO isotopologues and H$_{2}$CO (upper set of panels) are tracing the large-scale envelope and outflow structures, complex molecules (rows 2 to 7) are tracing a compact but often spatially-resolved emission centered on the maximum of the continuum emission. Some molecular lines are blended, see Table\,\ref{t:n1333-ir2a_linelist_s2} for further information. Contours show the 3$\sigma$, 5$\sigma$, and 8$\sigma$, and then 10$\sigma$ to 100$\sigma$ in 20$\sigma$ steps..}}
\label{fig:all_maps}
 \end{figure*}
 }

\end{document}